\documentstyle[12pt,epsf]{article}
\begin{document}
\hoffset=-1.2cm
\hsize=16cm
\vsize=24cm%\documentstyle[a4wide,12pt]{article}
\begin{flushright}
DTP-95/92\\
IC/95/353\\
October, 1995\\
\end{flushright}
\vspace*{1.5cm}
{\Large{\centerline{\bf {CONSTRAINT ON THE QED VERTEX FROM}}}}
\vskip 2mm
{\Large{\centerline{\bf {THE MASS ANOMALOUS DIMENSION $\gamma_m\,=\,1$}}}}
\vskip 2cm
\baselineskip=7mm
{\centerline{{\bf{A. Bashir $^{{\dagger}\,,\;{\dagger\dagger}}$ and
M.R. Pennington $^{\dagger}$}}}}
\vskip 1cm
{\centerline{Centre for Particle Theory,$^{\dagger}$}}
{\centerline{University of Durham}}
{\centerline{Durham DH1 3LE, U.K.}}
\vskip 5mm
{\centerline{International Centre for
Theoretical Physics,$^{{\dagger \dagger}}$}}
{\centerline{P.O. Box 586}}
{\centerline{34100 Trieste, Italy}}
\vskip 2cm
{\centerline {ABSTRACT}}
{\noindent We discuss the structure of the non-perturbative
fermion-boson vertex in quenched QED. We show that it is possible
to construct  a vertex which not only ensures that the
fermion propagator is multiplicatively
renormalizable, obeys the appropriate Ward-Takahashi identity,
reproduces perturbation theory for weak couplings and
guarantees that the critical coupling at which the mass is dynamically
generated is gauge independent
but also makes sure that the value for the anomalous
dimension  for the mass function is strictly 1, as Holdom and Mahanta have
 proposed.}
\vfil\eject
\vskip 2cm
\noindent
\baselineskip=8mm
\parskip=2.mm

In a recent paper~\cite{Mike}, we presented a mechanism for constructing
an effective non-perturbative vertex in quenched QED which incorporates
some of the key features required for a gauge theory. It ensures the
fermion propagator is multiplicatively renormalizable, the Ward-Takahashi
identity relating the fermion propagator to the fermion-boson vertex
is satisfied, reproduces perturbation theory for low values of the coupling
and yields a strictly gauge independent critical coupling for dynamical
mass generation.
This construction builds on the results
of Dong et al.~\cite{Dong}. The non-perturbative vertex is
written in terms of two unknown functions $W_{1}$ and $W_2$ which obey
certain conditions, Eqs.~(28,33,46,59) of~\cite{Mike}. With the fermion
propagator of momentum $p$ given by
\vspace{2mm}
\begin{eqnarray}
    S_{F}(p)&=& \frac{F(p^2)}{\not \! p- {\cal M}(p^2)}\qquad ,
\end{eqnarray}
the function
$W_{1}$ corresponds to the
equation for the fermion wavefunction renormalization $F(p^2)$, Eq.~(12),
while
$W_{2}$ is related to the mass function ${\cal M}(p^2)$, Eq.~(13) of
\cite{Mike}. One of the assumptions made in that work was
that the transverse vertex defined by Eqs.~(9,10) of~\cite{Mike} vanishes
in the Landau gauge as it does in the leading logarithm approximation.
Although this assumption does not enter the discussion of $W_{1}$,
the conditions for $W_{2}$, Eq.~(46,59) of~\cite{Mike},  crucially
depend on it. This issue is intimately
related to the value of the anomalous dimension for the fermion mass
function, $\gamma_m$. In the quenched theory the ultraviolet behaviour of
${\cal M}(p^2)$ can be expressed as
\vspace{2mm}
\begin{eqnarray}
 {\cal M}(p^2)\,\sim \, (p^2)^{\gamma_m/2-1} \label{defM}
\end{eqnarray}
in the deep Euclidean region. At criticality, where there is only one
momentum scale, $\Lambda$ the ultraviolet cutoff~\cite{Miransky}, the mass
function
behaves as in Eq.~(\ref{defM}) at all momenta.  If the aforementioned
assumption  about the vanishing Landau gauge transverse vertex holds true,
then $\gamma_m\,=\,1.058$.

 However,
Holdom~\cite{Holdom}, followed by Mahanta~\cite{Mahanta}, using arguments
based on  the Cornwall-Jackiw-Tomboulis
(CJT) effective potential technique has shown that $\gamma_m$ is strictly
equal to 1 regardless of the choice of the vertex. If this were true,
this would suggest
that there is a necessary piece in the transverse part of the effective
vertex which does not vanish in the Landau gauge.
A recent perturbative calculation of the transverse vertex in an arbitrary
covariant gauge, performed by K{\i}z{\i}lers\"u et al.~\cite{Ayse},
reveals that the
transverse part of the actual vertex {\em does not} vanish in the Landau
gauge.
This fact may possibly favour Holdom's conclusions. It may well be that
the non-zero transverse piece in the Landau gauge restores the
simplicity of the result which is the characteristic of the bare
vertex, spoiled by an additional term introduced in the longitudinal
vertex proposed by Ball and Chiu~\cite{BC}. In this comment, we show that
one could construct a vertex which ensures that the gauge invariant
chiral symmetry breaking takes place with $\gamma_m=1$ instead of
$\gamma_m=1.058$.
This vertex again involves the same function $W_1$, but instead of $W_2$,
we obtain a function $V_2$ which obeys slightly different conditions
compared to Eqs.~(46,59) of~\cite{Mike}.
We use the same definitions and notations as in [1] unless mentioned
otherwise.

Firstly, we recall that if the equation for $F(p^2)$ is to have a solution
that is
 multiplicatively renormalizable, then it must behave as
 \begin{equation}
 F(p^2)\,=\,\left({p^2 \over{\Lambda^2}}\right)^{\nu}\quad ,
 \label{FMR}
 \end{equation}
 where $\nu\,=\,\alpha\xi/4\pi$ in keeping
 with the Landau-Khalatnikov transformation~\cite{LK}.

It is then well known that in the case of the bare vertex, the mass
function
obeys the following linearized equation in Euclidean space
in the Landau gauge, where $F(p^2)\,=\,1$~:
\vspace {2mm}
\begin{eqnarray}
{\cal M}(p^2) &=& \frac{3 \alpha}{4\pi}\
                          \int_{0}^{\Lambda^2}
                          \frac{dk^2}{k^2} \, {\cal M}(k^2) \,
                           \left[ \frac{k^2}{p^2} \theta (p^2-k^2)
                          +  \theta (k^2-p^2) \right] \qquad.
\label{4.2bareM}
\end{eqnarray}
This equation has the multiplicatively renormalizable solution
of the form of  Eq.~(\ref{defM}), with
\begin{eqnarray}
\gamma_m = {1} \pm \sqrt {1- \frac{\alpha}{\alpha_{c}}}
\quad, \label{eq:shalf}
\end{eqnarray}
where $\alpha_c\,=\,\pi/3$. When
$\alpha\,=\,\alpha_c$, $\gamma_m=1$. (Note that in~\cite{Mike} the
exponent in Eq.~(\ref{defM}) is called $-s$, so
that $\gamma_m\,=\,2(1-s)$.)
In order that Eq.~(34) of~\cite{Mike} is identical to Eq.
(\ref{4.2bareM}) for all values of
the covariant gauge parameter, $\xi$, the following must hold true~:
\newpage
\vspace {2mm}
\begin{eqnarray}
 \nonumber            && \frac{\xi}{3} \int_{0}^{p^2}
                          \frac{dk^2}{p^2}\ {\cal M}(k^2)
                          \frac{F(k^2)}{F(p^2)}\\ \nonumber
                     &=&
                     - \vspace{5 mm} \int_{0}^{p^2} \frac{dk^2}{p^2}
                          \frac{1}{2(k^2-p^2)}
                       \Bigg[  p^2 {\cal M}(k^2)
                         \left( 1- \frac{F(k^2)}{F(p^2)}\
                         \right)
                   - k^2 \left( {\cal M}(k^2) -  {\cal
                         M}(p^2) \frac{F(k^2)}{F(p^2)} \right)  \Bigg]
                         \\  \nonumber
                   & & - \vspace{5 mm} \int_{p^2}^{\Lambda^2}
                          \frac{dk^2}{k^2}
                          \frac{1}{2(k^2-p^2)}
                       \Bigg[ k^2 {\cal M}(k^2)
                         \left( 1- \frac{F(k^2)}{F(p^2)}\
                         \right)
                   - p^2 \left( {\cal M}(k^2) -  {\cal
                         M}(p^2) \frac{F(k^2)}{F(p^2)} \right) \Bigg]
                       \\  \nonumber
                   & & + \vspace{5 mm} \int_{0}^{p^2}
                         \frac{dk^2}{p^2}
                         {\cal M}(k^2) F(k^2) \Bigg[ \frac{k^2}{6}
                         (k^2-3p^2)\,\tau_{2}(k^2,p^2)
                         +\,  p^2 \,\tau_{3}(k^2,p^2)
                        \, +\,  (k^2-p^2)\,\tau_{6}(k^2,p^2)
                             \Bigg] \\ \nonumber
                   & & + \vspace{5 mm} \int_{p^2}^{\Lambda^2}
                         \frac{dk^2}{k^2}
                       {\cal M}(k^2) F(k^2) \Bigg[  \frac{p^2}{6}
                         (p^2-3k^2)\, \tau_{2}(k^2,p^2)
                   +\, k^2 \,\tau_{3}(k^2,p^2)
                         \,+ \, (k^2-p^2)\, \tau_{6}(k^2,p^2)  \Bigg] \; ,
\\    \label{4.2inter1}
\end{eqnarray}
where the $\tau_i$, $i=2,3,6,8$ ($\tau_8$ only occurs in the
analogous equation for $F(p^2)$) are the coefficients of the transverse
vertex in the basis of Ball and Chiu~\cite{BC}.
Demanding that a chirally symmetric solution should be possible
when the bare mass is zero  is most easily accomplished
if only those transverse vectors with odd numbers of
gamma matrices contribute to the transverse vertex.
That is why, $\tau_i$, $i=1,4,5,7$, have dropped out.

Introducing in Eq.~(\ref{4.2inter1}) the variable $x$, where,
for $ 0 \le k^{2} < p^{2} $, $x=k^{2}/p^{2}$, and for $ p^{2} \le k^{2} <
\Lambda^2 $,
$x=p^{2}/k^{2}$, we can now retrace the steps carried out in obtaining
Eq.~(46) in~\cite{Mike}, starting
from Eq.~(41). This will lead us to the compact equation
\vspace{2mm}
\begin{eqnarray}
       \int_{0}^{1} \frac{dx}{\sqrt{x}} \;V_2(x)=0 \qquad ,
\label{4.2V2}
\end{eqnarray}
where using Eq.~(\ref{FMR})
\vspace{2mm}
\begin{eqnarray}
\nonumber    V_2(x)&=& \xi\ x^{\nu } + \frac{3}{2}\
          \left[  \frac{x^{-\nu}-x^{\nu}}{x-1} \right] - \frac{3x}{2}
         \left[   \frac{x^{-(
\nu+ \frac{1}{2})} - x^{(\nu+ \frac{1}{2})}}{x-1} \right]  \\
&& - x^{\nu} \left[ g_1(x) + g_2(x) \right] - x^{-\nu} \left[ g_1(1/x) -
g_2(1/x) \right]  \;.  \label{V2W2}
\end{eqnarray}
The functions $g_1(x)$ and $g_2(x)$ are as defined in~\cite{Mike}.
Here, the function $V_2$ is the counterpart of $W_2$ in~\cite{Mike}.
Eqs.~(\ref{4.2V2},\ref{V2W2}) should be compared with Eqs.~(46,47) of
\cite{Mike}. The condition Eq.~(\ref{4.2V2}) ensures that $\gamma_m=1$
in any covariant gauge, just
as condition Eq.~(46) ensures $\gamma_m=1.058$. As expected,
unlike $W_2$, $V_2$ does not vanish in the
Landau gauge. Instead, it is
\begin{eqnarray}
V_2(x)=  \frac{3\sqrt{x}}{2}  \;  - \; 2 \left[ g_1(x) +
g_2(x) \right]   \quad.
\end{eqnarray}
 Eq.~(\ref{V2W2}) can be inverted to evaluate the expressions for $\tau_i$
in terms of $V_2$. We shall not give the expression for
$\tau_6$ as it is solely related to the equation for the wavefunction
renormalization, Eq.~(14) of~\cite{Mike}, and hence remains completely
independent of the value of $\gamma_m$. Repeating the same steps as
in~\cite{Mike},
we obtain
%\newpage
\begin{eqnarray}
\nonumber \tau_2(k^2,p^2)&=& -6 \,\frac{ \tau_6(k^2,p^2)}{
          (k^2-p^2)} \\  \nonumber
       & & +    \frac{1}{(k^2-p^2)^2}
               \frac{1}{\left[F(k^2)+F(p^2)\right]} \left\{
               3 (k^2+p^2)
          R_2(k^2,p^2) +  2\xi Q_2(k^2,p^2) \right\} \\ \nonumber
       & &-\frac{1}{(k^2-p^2)^2}\  \frac{1}{\left[F(k^2)+F(p^2)\right]}\
           \Bigg[ V_2\left( \frac{k^2}{p^2}\ \right) +  V_2\left(
            \frac{p^2}{k^2}\ \right) \Bigg]  \\
       & &-\frac{k^2+p^2}{(k^2-p^2)^3}\
           \frac{1}{\left[F(k^2)+F(p^2)\right]}\
           \Bigg[ V_2\left( \frac{k^2}{p^2}\ \right) -  V_2\left(
            \frac{p^2}{k^2}\ \right) \Bigg] \, ,   \label{tau2}
\end{eqnarray}
where both $Q_2(k^2,p^2)$ and $R_2(k^2,p^2)$ are symmetric functions of $k$
and $p$~:
\vspace{2mm}
\begin{eqnarray}
\nonumber
  Q_2(k^2,p^2)&=& \frac{1}{k^2-p^2} \Bigg[
              k^2 \, \frac{F(k^2)}{F(p^2)}
             - p^2 \,
              \frac{F(p^2)}{F(k^2)} \Bigg]  \\
  R_2(k^2,p^2)&=& \frac{1}{k^2-p^2} \Bigg[
              \frac{F(k^2){\cal M}(p^2)}{F(p^2){\cal M}(k^2) }
             - \frac{F(p^2){\cal M}(k^2)}{F(k^2){\cal M}(p^2)} \Bigg] \;,
\end{eqnarray}
\vspace {2mm}
\begin{eqnarray}
\nonumber \tau_3(k^2,p^2)&=&  - \frac{k^2+p^2}{
          k^2-p^2}\, \tau_6(k^2,p^2) \\ \nonumber
          & &+ \frac{1}{(k^2-p^2)^2} \frac{1}{ \left[F(k^2)+F(p^2)
          \right]} \left\{  2 k^2 p^2
          R_2(k^2,p^2) -  \frac{\xi}{3} Q_3(k^2,p^2) \right\} \\ \nonumber
          & &- \frac{1}{6} \frac{k^2+p^2}{(k^2-p^2)^2}\
          \frac{1}{ \left[F(k^2)+F(p^2)\right]}\ \Bigg[ V_2\left(
           \frac{k^2}{p^2}\ \right) +
          V_2\left( \frac{p^2}{k^2}\ \right) \Bigg]  \\ \nonumber
          & & +\frac{1}{6} \frac{k^4+p^4-6k^2p^2}{(k^2-p^2)^3}\
          \frac{1}{ \left[F(k^2)+F(p^2)\right] }\  \Bigg[ V_2\left(
          \frac{k^2}{p^2}\ \right) -
          V_2\left( \frac{p^2}{k^2}\ \right) \Bigg] \, , \\
\end{eqnarray}
where
\vspace{2mm}
\begin{eqnarray}
 Q_3(k^2,p^2)&=& \frac{1}{k^2-p^2} \left[ p^2 (p^2-3k^2)
\frac{F(k^2)}{F(p^2)} - k^2 (k^2-3 p^2) \frac{F(p^2)}{F(k^2)} \right]
 \quad.
\end{eqnarray}
and
\vspace{2mm}
\newpage
\begin{eqnarray}
\nonumber
          \tau_8(k^2,p^2)&=&  -2\, \frac{k^2+p^2}{
          k^2-p^2} \,\tau_6(k^2,p^2) + \overline{\tau}(k^2,p^2) \\
          \nonumber
          & &+ \frac{1}{(k^2-p^2)^2}
             \frac{1}{\left[F(k^2)+F(p^2)\right]} \\ \nonumber
          & & \hspace{10mm} \left\{
           \frac{1}{2} (3k^4+3p^4+2k^2p^2)
           R_2(k^2,p^2) +  \frac{\xi}{3} Q_8(k^2,p^2) \right\} \\ \nonumber
          & &- \frac{1}{3}\ \frac{k^2+p^2}{(k^2-p^2)^2}\
            \frac{1}{\left[F(k^2)+F(p^2)\right] }\
           \Bigg[ V_2\left( \frac{k^2}{p^2}\ \right) +  V_2\left(
            \frac{p^2}{k^2}\ \right) \Bigg]    \\ \nonumber
          & &- \frac{2}{3}\ \frac{k^4+p^4}{(k^2-p^2)^3}\
            \frac{1}{\left[F(k^2)+F(p^2)\right] }\
           \Bigg[ V_2\left( \frac{k^2}{p^2}\ \right) -  V_2\left(
            \frac{p^2}{k^2}\ \right) \Bigg]  \, , \\
\end{eqnarray}
where
\vspace{2mm}
\begin{eqnarray}
 Q_8(k^2,p^2)&=&\frac{1}{(k^2-p^2)} \left[ (3k^4 + p^4)
                \frac{F(k^2)}{F(p^2)} - (k^4+3p^4) \frac{F(p^2)}{F(k^2)}
                \right]  \; \label{Q8}
\end{eqnarray}
and $\overline{\tau}(k^2,p^2)$ is defined by Eq.~(17) of~\cite{Mike}.
Note that all the momenta above are in Euclidean space.
Therefore, appropriate changes of sign have to be made in order
to get the expressions for $\tau_i$ in the Minkowski space
to construct the transverse vertex using the basis vectors of
Ball and Chiu~\cite{BC}.
Eqs.~(\ref{tau2}--\ref{Q8}) should be compared with Eqs.~(52--58)
of~\cite{Mike}.
It is here that we note the restoration of simplicity.
In contrast with $q_i(k^2,p^2)$, $Q_i(k^2,p^2)$ do not have any dependence
on the mass functon ${\cal M}(p^2)$ at all.
Moreover, the explicit appearance of the mass term in $\tau_2$, $\tau_3$
and
$\tau_8$ in the present case is only through the factor
\vspace{2mm}
\begin{eqnarray*}
  \frac{1}{k^2-p^2} \Bigg[
              \frac{F(k^2){\cal M}(p^2)}{F(p^2){\cal M}(k^2) }
             - \frac{F(p^2){\cal M}(k^2)}{F(k^2){\cal M}(p^2)} \Bigg] \;,
\end{eqnarray*}
unlike the case with $\gamma_m=1.058$, where $r_2(k^2,p^2)$, $q_2(k^2,p^2)$,
$q_3(k^2,p^2)$ and $q_8(k^2,p^2)$ all carry the dependence on the  mass
function
${\cal M}(p^2)$ through different and more complicated terms.

Imposing the condition that the vertex and its components should be
free of kinematic singularities now implies,
\vspace{2mm}
\begin{eqnarray}
             V_{2}(1)+2V'_{2}(1)= 2\nu (\xi+3)+ (\xi+6-3\gamma_m) \qquad ,
\label{4.3deriv}
\end{eqnarray}
which replaces Eq.~(59) of~\cite{Mike} and at the critical coupling, it
reduces to  $V_{2}(1)+2V'_{2}(1)=  (\xi+3) (2\nu+1)$.
The transverse vertex now has the correct lowest order
perturbative limit, viz. $\Gamma_T^{\mu} = {\cal O}(\alpha)$, provided
\vspace{2mm}
\begin{eqnarray}
 V_{2}\left(k^2/p^2\right)\,=\,\xi\ \frac{F(k^2)}{F(p^2)}\ +
                \frac{3}{2} \left[ \frac{F(k^2){\cal
         M}(p^2)}{F(p^2){\cal M}(k^2)}\, - \, \frac{F(p^2){\cal
         M}(k^2)}{F(k^2){\cal M}(p^2)} \right] + \,{\cal O}(\alpha)
\quad.  \label{4.4pert}
\end{eqnarray}
Note that Eq.~(\ref{4.2V2}) is only true at the
bifurcation point just as Eq.~(46) in~\cite{Mike} whose exact form for all
$\alpha$ might be suggested by Eq.~(\ref{eq:shalf}) to be
\vspace{2mm}
\begin{eqnarray}
       \int_{0}^{1}  {dx\over{\sqrt{x}}} \; V_2(x)\,\approx\, 2 \, \xi \,
       \sqrt{1-{\alpha\over{\alpha_c}}}\quad.
\end{eqnarray}
in order to agree with  both the $\alpha = 0$ and
$\alpha = \alpha_c$ limits, Eqs.~(\ref{4.4pert},\ref{4.2V2}).

We have been able to show that there is no technical difference between
the mechanism of constructing the transverse vertex for the case
$\gamma_m=1$ and
$\gamma_m=1.058$.
On comparing Eq.~(47) of~\cite{Mike} and Eq.
(\ref{V2W2}), we can see that the main difference between $V_2$ and $W_2$
is that $V_2$ has an additional piece coming from the longitudinal
part of the vertex. As a result of
this difference,  $V_2$  does not vanish in the Landau gauge in
contrast to $W_2$. It is this that ensures at criticality that
 the anomalous dimension
$\gamma_m=1$ identically in all covariant gauges in keeping with the
results
of~\cite{Holdom,Mahanta}.

\vskip 1cm

\noindent{\Large{\bf Acknowledgements}}

   \vskip 5mm
 AB wishes to thank the Government of Pakistan for a research studentship
 and International Centre for Theoretical Physics (ICTP), Trieste, for
 their hospitality and support during his visit to the Centre.

\vfil\eject
\hsize=16.5cm
\baselineskip=6mm

\end{document}